\documentclass[reprint,twocolumn,floats, floatfix,superscriptaddress,amsmath,amssymb,aps]{revtex4-1}

\usepackage{graphicx}
\usepackage{dcolumn}
\usepackage{bm}
\usepackage{color}

\begin{document}


\title{Fluidization of epithelial sheets by active cell rearrangements}
\author{Matej Krajnc}
\email{mkrajnc@princeton.edu}
\affiliation{Lewis-Sigler Institute for Integrative Genomics, Princeton University, Washington Road, Princeton, USA}
\affiliation{Jo\v zef Stefan Institute, Jamova 39, SI-1000, Ljubljana, Slovenia}
\author{Sabyasachi Dasgupta}
\affiliation{Department of Physics, University of Toronto, 60 St.~George St., Toronto, ON M5S 1A7, Canada}
\affiliation{Mechanobiology Institute, National University of Singapore, 5A Engineering Drive 1, Singapore 117411, Singapore}
\author{Primo\v z Ziherl}
\affiliation{Jo\v zef Stefan Institute, Jamova 39, SI-1000, Ljubljana, Slovenia}
\affiliation{Faculty of Mathematics and Physics, University of Ljubljana, Jadranska 19, SI-1000 Ljubljana, Slovenia}
\author{Jacques Prost}
\affiliation{Mechanobiology Institute, National University of Singapore, 5A Engineering Drive 1, Singapore 117411, Singapore}
\affiliation{Laboratoire Physico Chimie Curie, Institut Curie, PSL Research University, CNRS UMR168, 75005 Paris, France}


\begin{abstract}

We theoretically explore fluidization of epithelial tissues by active T1 neighbor exchanges. We show that the geometry of cell-cell junctions encodes important information about the local features of the energy landscape, which we support by an elastic theory of T1 transformations. Using a 3D vertex model, we show that the degree of active noise driving forced cell rearrangements governs the stress-relaxation time-scale of the tissue. We study tissue response to in-plane shear at different time scales. At short time, the tissue behaves as a solid, whereas its long-time fluid behavior can be associated with an effective viscosity which scales with the rate of active T1 transformations. Furthermore, we develop a coarse-grained theory, where we treat the tissue as an active fluid and confirm the results of the vertex model. The impact of cell rearrangements on tissue shape is illustrated by studying axial compression of an epithelial tube.
\end{abstract}

\maketitle

\section{Introduction} 
In the earliest stages of animal development, many embryos consist of an epithelial sheet-like tissue which then deforms so as to form the various body parts, the main morphogenetic modes relying on buckling, invagination, spreading, and in-plane migration. The three-dimensional (3D) tissue shapes resulting from these processes depend very much on the ability of cells to move within the tissue due to local stresses imposed by their neighbors~\cite{rauzi15,etournay15}, which is intrinsically associated with the energy barrier for cell rearrangement~\cite{bi14,bi15}. 

This motion is reminiscent of glassy dynamics~\cite{angelini11,shoetz13} known from colloidal suspensions and granular materials~\cite{berthier11}. One of its key features is the transition from the fluid-like state to a state characterized by a finite shear modulus. For example, in flat non-proliferating epithelia described by the 2D area- and perimeter-elasticity (APE) model~\cite{farhadifar07,hufnagel07}, the energy barriers for cell rearrangements vanish at strong cell-cell adhesion or at weak contractility of the actomyosin belt~\cite{bi15}. In turn, intrinsically solid tissues can actively reorganize by oriented cell intercalations~\cite{irvine94,keller06,rauzi10,popovic17} or they can be fluidized by cell proliferation~\cite{farhadifar07,ranft10,matozfernandez17a,matozfernandez17b} and active cell motility~\cite{bi16,yang17,barton17}. These interesting findings raise the question whether non-proliferating solid tissues in which cells do not interact strongly with the substrate too can exhibit fluid-like behavior governed by, e.g., active remodeling of cell-cell junctions, and if yes, how such activity affects the overall 3D tissue shape. This could be especially important in the tissues of early embryos, which undergo extensive 3D shape transformations, driven almost exclusively by intra- and inter-cellular forces. Another important issue concerns the reduced dimensionality of the 2D models, which provides valuable insight into the in-plane structure~\cite{farhadifar07,hufnagel07,hocevar09,staple10} and out-of-plane deformations of the tissue~\cite{hocevarbrezavscek12,krajnc15,storgel16,osterfield14,fletcher14,murisic15,monier15} yet leaves a lingering question of whether the approximations used are really justified.

\begin{figure}[t!]
\centerline{\includegraphics{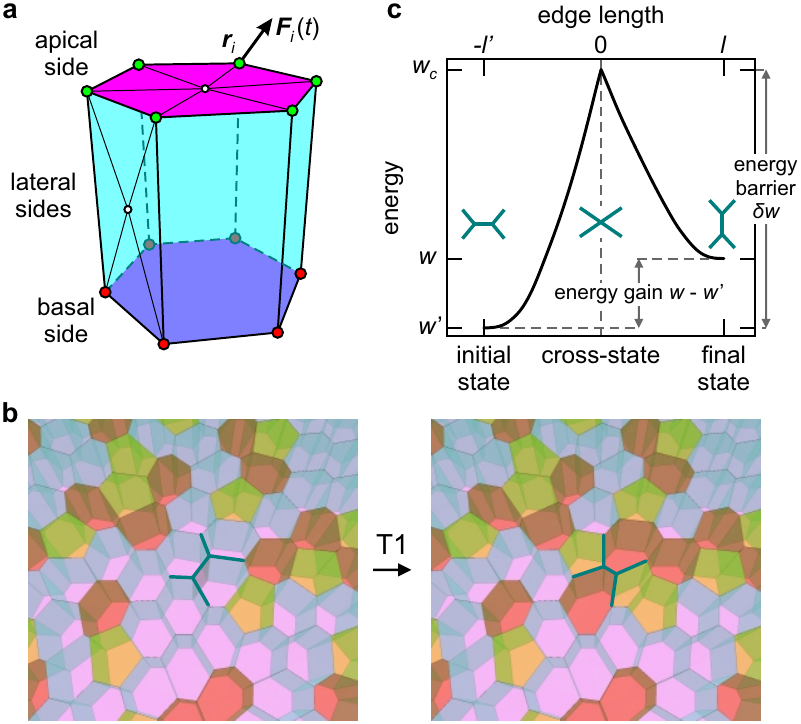}}
\caption{\label{fig1}(a)~Schematic of the 3D vertex model. Basal and apical vertices (red and green circles, respectively) move during simulation depending on force $\boldsymbol F_i$. (b)~Snapshots of a tissue showing a T1  transformation, with the apical sides of 5-, 6-, and 7-coordinated cells in semitransparent yellow, white, and brown, respectively. Edges undergoing the transformation are highlighted. (c)~Schematic of tissue energy during a T1 transformation, showing energy barrier $\delta w=w_c-w'$ and energy gain $w-w'$.}
\end{figure}

To address these problems, we develop a theoretical model of the epithelial tissue, in which cells undergo active neighbor exchanges due to random non-directional contractions of cell-cell junctions induced by, e.g., fluctuations of the junctional myosin. Combining this phenomenon with the recently proposed 3D vertex model~(Fig.~\ref{fig1}a;~\cite{okuda15,bielmeier16,misra16,alt17}) allows us to construct a detailed computational representation of the tissue, which is then used to investigate the complex relationship between cell-scale activity and 3D tissue architecture. In particular, we show that the geometry of the network of cell-cell junctions reveals important features of the energy landscape associated with the tissue. We use this result to study how local junctional activity controls the viscoelastic properties of the tissue. We find a simple scaling relation between the effective viscosity and the rate of active T1 transformations which is further supported by a coarse-grained theory. Finally, we demonstrate how these active processes significantly affect the global tissue shape by studying the relaxation dynamics of an epithelial tube under axial compression. 

\section{The model}
We describe epithelial sheets as aggregates of cells under tension. Due to differential adhesion and cortex contractility at the cell-lumen, cell-cell, and cell-substrate interfaces, the effective tensions on apical, basal, and lateral cell sides ($\Gamma_a$, $\Gamma_b$, and $\Gamma_l$, respectively) are all different~\cite{glazier93,hannezo14}. For the sake of simplicity, we consider tissues where each of these tensions is the same in all cells and we choose $\Gamma_a=\Gamma_b=\Gamma$. The shape of each cell is stabilized by an incompressible interior, which imposes a fixed-volume constraint $V_i=V$. We choose $V^{1/3}$ and $\Gamma_l$ as units of length and tension, respectively, and thus the dimensionless energy measured in units of $\Gamma_l V^{2/3}$ reads
\begin{equation}
	\label{eq1}
	w=\sum_{i=1}^{\mathcal C} \left [\gamma\left ( a_a^{(i)}+ a_b^{(i)}\right )+\frac{1}{2}a_l^{(i)}\right ]\>,
\end{equation}
where $\mathcal C$ is the number of cells, $a_a^{(i)}$, $a_b^{(i)}$, and $a_l^{(i)}$ are the dimensionless areas of the apical, basal, and lateral sides of cell $i$, respectively, and $\gamma=\Gamma/\Gamma_l$ is the dimensionless apical and basal tension; the lateral sides are shared by two cells, hence the factor of 1/2. The dynamics of cells is given implicitly by the friction-dominated equation of motion for the vertices: $\textup d\boldsymbol r_i(t)/\textup d t=\mu_i\boldsymbol F_i(t).$ Here $\boldsymbol r_i(t)$ is the position of vertex $i$ (Fig.~\ref{fig1}a), $\mu_i=\mu_0$ is its mobility assumed to be the same in all vertices, and $\boldsymbol F_i(t)=-\nabla_i w(\boldsymbol r_1, \boldsymbol r_2, \boldsymbol r_3,\ldots)$ is the force on vertex $i$; $\nabla_i=\left (\partial/\partial x_i,\partial/\partial y_i,\partial/\partial z_i\right )$. This equation is solved using a forward finite-difference scheme, the characteristic time scale being $\tau=(\mu_0\Gamma_l)^{-1}$. For details of implementation, see Sec.~I of Supplemental Material~\cite{SuppInf}. 
\section{Energies of T1 transformation}
The energy landscape of the epithelium is very complex, with each local minimum corresponding to a different in-plane arrangement of cells. The above model dynamics drives the system towards a given minimum, whereas jumping between minima entails topological changes facilitated by T1 transformations~(Fig.~\ref{fig1}b), cell divisions, and cell extrusions~\cite{etournay15,rauzi10,marinari12,wyatt16,rosenblatt17}. We focus on T1 transformations which are crucial for cell rearrangements and involve four cells around an edge. This edge first shrinks to form a cross-state with a four-way vertex which then dissociates into two three-way vertices, creating a new edge between initially non-neighboring cells~(Fig.~\ref{fig1}c). The initial and the cross-state are separated by an energy barrier $\delta w$ which remains finite at any $\gamma$ unlike in the APE model where it vanishes beyond the rigidity transition~\cite{farhadifar07,bi15}. Thus our tissue is intrinsically solid and any cell rearrangement requires an injection of energy via forced topological transformations. In living tissues, these injections come from active contractions of cell cortex driven by the junctional myosin~\cite{curran17,prost15}, the time scale for T1 transformations typically being on the order of minutes to hours~\cite{wyat16}.

To quantify the impact of active cellular rearrangements, we first study how the two energies associated with a T1 transformation (energy barrier $\delta w$ and energy gain $w-w'$ defined as the energy difference between the final and the initial state; Fig.~\ref{fig1}c) depend on network geometry. 
\begin{figure*}[htb!]
	\begin{center}
	\includegraphics[]{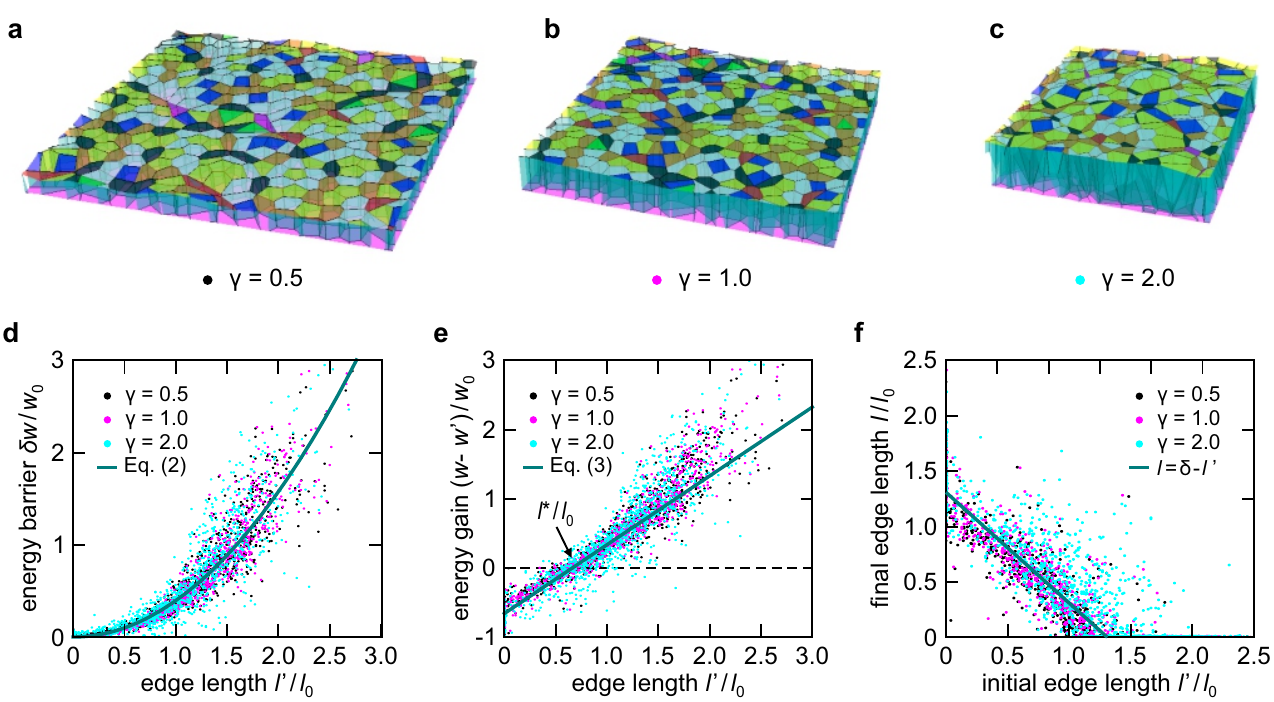}
	\caption{\label{F_S4}
(a-c)~Model tissue samples at $\gamma=0.5,\>1.0,\>\textup{and}\>\>2.0$ corresponding to a squamous, a short columnar, and a tall columnar tissue, respectively. Note that basal edges can have considerably different lengths than their apical counterparts. This effect is expected due to high degree of disorder, which gives rise to large local variations of the preferred Gaussian curvature---an effect especially prominent in the columnar tissue, e.g., panel c. Energy barrier $\delta w/w_0$ vs.~edge length $l'/l_0$~(d) and energy gain $(w-w')/w_0$ vs.~edge length $l'/l_0$ (e) for $\gamma=0.5,\>1,$ and~$2$; dots are simulation data and the solid lines are the parabola~[Eq.~(\ref{Eqbarrier})] and the linear function~[Eq.~(\ref{Eqgain})]. On average, T1 transformations are preferred below the threshold value $l^*/l_0\approx 0.66$~(e). (f)~Relation between the equilibrium edge length before and after T1 transformation ($l'$ and $l$, respectively); dots are simulation data and the solid line is the linear function $l=\delta-l'$, where $\delta\approx1.25l_0$. }
	\end{center}
\end{figure*}
We create a 3D cell sheet with a random network of cell-cell junctions by randomly distributing $\mathcal C=340$ points in a 2D square box of size $\mathcal Ca_0$, where $a_0=(\sqrt{3}/2)^{1/3}(2\gamma)^{-2/3}$ is the optimal surface area of the cell base~(Appendix~\ref{appA}). We then construct 2D Voronoi partitions around this set of points and apply periodic boundary conditions on box walls. From the obtained 2D polygonal network we build a 3D tissue and evolve the structure in time so as to satisfy the fixed-volume constraints and to find the minimal-energy structure at the Voronoi-constructed network topology. 

We perform T1 transformation on all cell-cell contacts and measure $\delta w$ and $w-w'$ as functions of $l'=(l_a+l_b)/2$, the average rest length of apical and basal edge of the side in question, for $\gamma=0.5,1,$ and 2 corresponding to a squamous, a short columnar, and a tall columnar epithelium (Fig.~\ref{F_S4}a, b, and c, respectively). By rescaling the energy by $w_0=2\gamma^{1/3}/3^{5/6}$ and edge lengths by $l_0=3^{-2/3}\gamma^{-1/3}$~(Appendix~\ref{appB}), we can almost perfectly collapse the data for both $\delta w$ and $w-w'$~(Fig.~\ref{F_S4}d and e). In particular, we find
\begin{equation}
	\label{Eqbarrier}
	\delta w= 2\sqrt{3}\gamma R l'^2\>,
\end{equation}
and
\begin{equation}
	\label{Eqgain}
	w-w'=\frac{2\gamma^{2/3}Q}{3^{1/6}}\left (l'-l^*\right )\>,
\end{equation}
where the fitting parameters $R\approx 0.4$ and $Q\approx 1$, and the threshold length $l^*\approx0.66l_0=0.66\cdot 3^{-2/3}\gamma^{-1/3}$. Moreover, we find the same forms of $\delta w(l')$ and $(w-w')(l')$ in two other mechanical models---the 2D APE model and the 2D foam~(Appendix~\ref{appC})---which suggests that relations given by Eqs.~(\ref{Eqbarrier}) and (\ref{Eqgain}) might be generic. This has been recently confirmed in 2D tissue and soap-froth models \cite{sascha18}.

To better understand this, we next develop an elastic theory of T1 transformation. Here, the cross state~(Fig.~\ref{fig1}c) can be seen as an elastic deformation of the edge network, described by the shear modulus $G$ (measured in Sec.~\ref{secIV}). Elastic energy difference between the cross state and the initial state $\delta w=w_c-w'$ reads
\begin{equation}
	\delta w=\frac{G}{2}\int\textup dA\left [\left (\partial_xu_x-\partial_yu_y\right )^2+\left (\partial_xu_y+\partial_yu_x\right )^2\right ]\>,
\end{equation}
where $u_x$ and $u_y$ denote displacements along orthogonal directions. The deformation is obtained by applying a force $f_x$ at $x=l'/2$ and an opposite force ($-f_x$) at $x=-l'/2$. We neglect a possible residual anisotropy due to finite size of the system and for the sake of simplicity, we assume that the network is incompressible such that 
\begin{equation}
	\partial_xu_x+\partial_yu_y=0\>
\end{equation}
(keeping finite compressibility would not change the final result). Force balance imposes
\begin{equation}
	G\left (\partial_{xx}^2+\partial_{yy}^2\right )u_x-\partial_x p=f_x\delta_{x-l'/2}-f_x\delta_{x+l'/2}
\end{equation}
and
\begin{equation}
	G\left (\partial_{xx}^2+\partial_{yy}^2\right )u_y-\partial_y p=0\>,
\end{equation}
where $p$ is pressure and at the boundaries $u_x(\pm l'/2)=\mp l'/2$ and $u_y(\pm l'/2)=0$. The equations are solved in Fourier space. We obtain
\begin{equation}
	u_x\left (\frac{l'}{2}\right )=\frac{f_x}{8\pi G}\left [\psi-\textup{Ci}(q_cl')+\ln{(q_cl')}\right ]=-\frac{l'}{2}\>,
\end{equation}
where we introduced an integration cutoff wavenumber $q_c$. In turn, it implies 
\begin{equation}
	f_x=-\frac{4\pi G}{\psi-\textup{Ci}(q_cl')+\ln{(q_cl')}}l'\>.
\end{equation}
Here $\psi=0.5772156649...$ is the Euler-Mascheroni constant and $\textup{Ci}(x)=\psi+\ln x+\int_0^x\textup dt(\cos t-1)/t$. Now we can calculate the energy barrier $\delta w$ from its expression in Fourier space 
\begin{equation}
	\delta w=\frac{f_x^2}{2\pi^2G}\int\frac{\textup dq_x}{2\pi}\frac{\textup dq_y}{2\pi}\frac{q_y^2}{(q_x^2+q_y^2)^2}\sin^2\left (\frac{q_xl'}{2}\right )
\end{equation}
and obtain
\begin{equation}
	\label{eqBarrierSI}
	\delta w=\frac{2\pi G}{\psi-\textup{Ci}(q_cl')+\ln{(q_cl')}}l'^2\>.
\end{equation}
The force needs to vanish when $l'\to 0$, which implies $q_c\propto 1/l'$. Comparing Eq.~(\ref{eqBarrierSI}) with Eq.~(\ref{Eqbarrier})  gives $a_c=q_cl'\approx 9.877$, such that $\delta w\propto l'^2$, in agreement with the numerical results. 
\begin{figure*}[htb!]
	\begin{center}
	\includegraphics[]{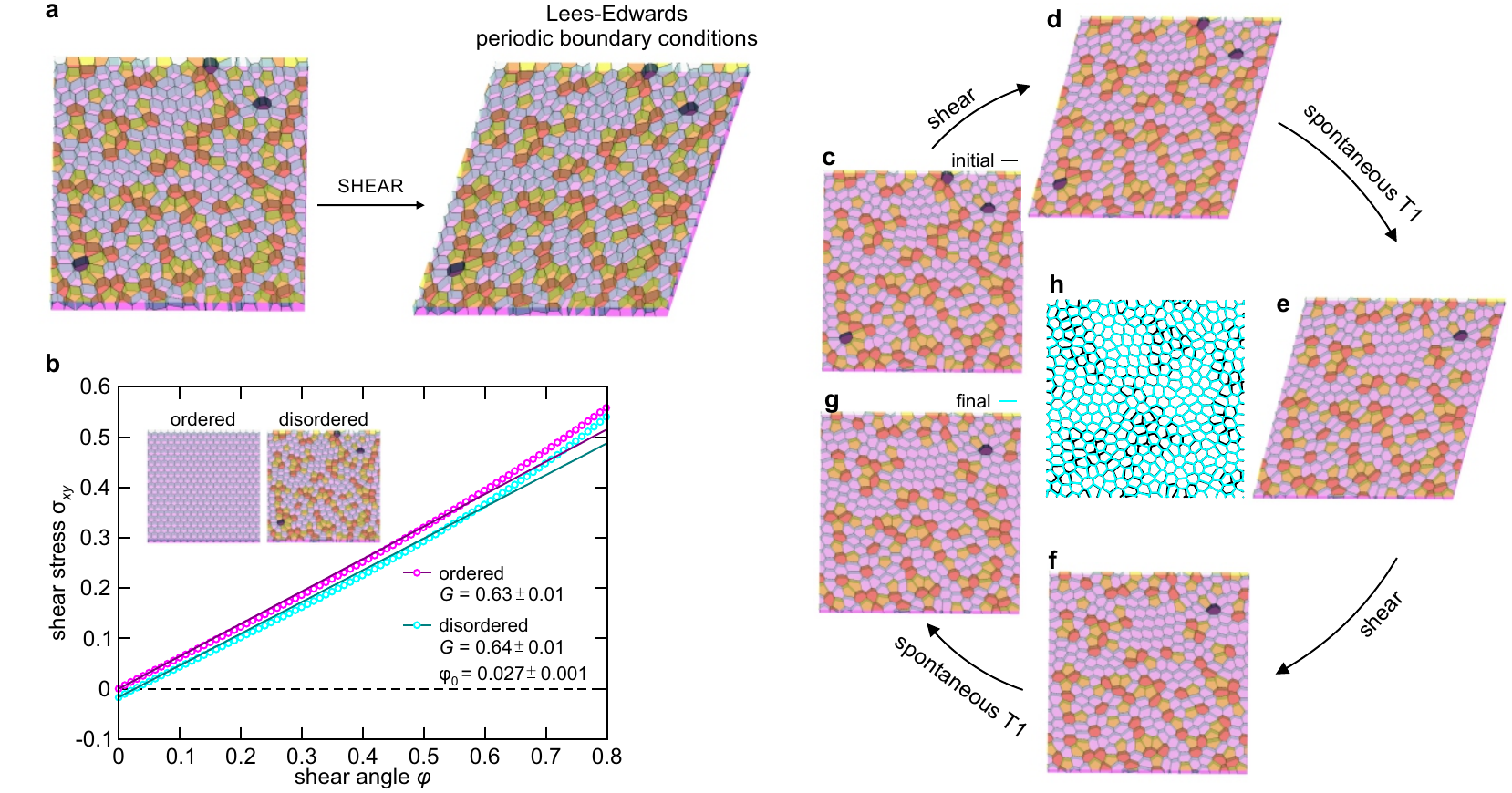}
	\caption{\label{F_S8}
	(a)~An example of a partially randomized flat 3D epithelial cell sheet sheared by the use of Lees-Edwards periodic boundary conditions. (b)~Shear stress $\sigma_{xy}$ as a function of shear strain $\varphi$ for an ordered cell packing (magenta circles) and a disordered cell packing (cyan circles) at $\gamma=1$. Shear modulus $G$ is extracted from the slope of the linear fit at $\varphi=0$. (c-g)~Snapshots of simulations in the direction of arrows show a plastic shear deformation of the tissue. On a short time scale, the tissue is shear-deformed~(c-d) and on a long time scale, it undergoes T1 transformations~(d-e). After applying a reverse shear deformation (e-f), the tissue deforms back towards the initial shape but the final state~(g) is not exactly identical to the initial state because of the irreversible T1 transformations that took place. In the central panel, the initial and the final edge networks (black and cyan edges, respectively) are compared~(h).}
	\end{center}
\end{figure*}

After the T1 transformation, there exist another equilibrium with energy $w$. If we apply a force dipole acting along the $y$-direction, we can bring the system to the same cross point with displacement along the $y$-direction. The same calculation as above relates $w_c$ with $w$ with a correction due to the fact that the T1 transformation introduced pairs of pentagons and heptagons:
\begin{equation}
	w_c-w=\frac{2\pi G}{\psi-\textup{Ci}(a_c)+\ln{(a_c)}}l^2+w_{5-7}\>.
\end{equation}
We find
\begin{equation}
	w-w'=\frac{2\pi G}{\psi-\textup{Ci}(a_c)+\ln{(a_c)}}(l'^2-l^2)-w_{5-7}\>,
\end{equation}
where the relation between $l$ and $l'$ is not known {\it a priori}. The simu\-lation results show that at lowest order, this relation is linear~(Fig.~\ref{F_S4}f): $l=\delta-l'$, such that, in agreement with simulations,
\begin{equation}
	\label{EnGainEqS}
	w-w'=\frac{4\pi G\delta}{\psi-\textup{Ci}(a_c)+\ln{(a_c)}}\left (l'-l^*\right )\>,
\end{equation}
where the threshold length
\begin{equation}
	l^*=\frac{\delta}{2}+\frac{w_{5-7}\left [\psi-\textup{Ci}(a_c)+\ln{(a_c)}\right ]}{4\pi G\delta}\>.
\end{equation}
Comparing Eqs.~(\ref{EnGainEqS}) and (\ref{Eqgain}) gives $\delta/l_0=1.25$, which again agrees with simulations~(Fig.~\ref{F_S4}f). These results further confirm that the relations given by Eq.~(\ref{Eqbarrier}) and Eq.~(\ref{Eqgain}) are not specific to our model.

%
%
%
%
\section{Active T1 transformations\label{secIV}}
The above results allow us to efficiently simulate random T1 transformations driven by junctional activity. To this end, we adapt the standard Metropolis algorithm such that the uphill active transformations on edges longer than $l^*$ take place with a finite probability independent of the rest length $l'$, whereas the probability of their downhill spontaneous counterparts on edges shorter than $l^*$ is 1 as usual. In particular, the probability for the former is given by $k_{\textup{T1}}\delta t/\mathcal E$, where $\mathcal E$ is the number of edges, $\delta t=0.001$ is the time step, and $k_{\textup{T1}}$ is the rate of active T1 transformations measured in units of $1/\tau$. Within the simplest non-equilibrium scheme, the imposed active noise is large compared to the barrier height and so the T1 rate does not explicitly depend the height of the energy barrier.

To quantify the viscoelastic response, we compute the stress caused by a sudden in-plane simple-shear strain in a slightly disordered $\gamma=1$ tissue with $\mathcal C=340$ cells and $\mathcal E=1020$ edges (Fig.~\ref{F_S8}a). In units of $\Gamma_lV^{-1/3}$, the $xy$ component of the stress reads
\begin{equation}
	\label{eq7}
	\sigma_{xy}=-\frac{1}{\mathcal C}\sum_{m\in\textup{lateral sides}} n_x^{(m)} n_y^{(m)}\>a_m\>,
\end{equation}
where $a_m$ is the area of lateral side $m$, whereas $n_x^{(m)}$ and $n_y^{(m)}$ are the components of its normal~\cite{batchelor70}. 

A short-time or small-deformation response to a simple-shear deformation is elastic. The time scale associated with the elastic behavior is much shorter than the time scale for T1 transformations $1/k_{\rm T1}$. Thus the tissue deforms at a fixed network topology and is able to relax back to the initial state when the reverse deformation is applied. In the linear regime, shear stress $\sigma_{xy}$ is proportional to shear strain $\varphi$:
\begin{equation}
	\label{E_S36}
	\sigma_{xy}=G\varphi\>,
\end{equation}
where the proportionality coefficient is the shear modulus $G$. We extract $G$ from simulations of a quasi-static shear deformation of the $\gamma=1$ tissue by calculating the slope of the $\sigma_{xy}(\varphi)$ curve: $G=\left.(\textup d\sigma_{xy}/\textup d\varphi)\right\vert_{\varphi=0}$. Figure~\ref{F_S8}b shows $\sigma_{xy}$ as a function of the shear angle~$\varphi$. In both ordered and disordered case, the linear regime extends to $\varphi\approx 0.5$. As expected, the shear modulus $G$ does not depend on the in-plane  structure of the tissue. However, topology does influence the preferred shear angle $\varphi_0$ which is equal to zero in the ordered tissue and generally non-zero albeit very small in disordered tissues.

For moderate to large deformations at short times, the tissue undergoes a plastic deformation. This happens due to spontaneous T1 transformations, which change the topology of the tissue so that after strain is removed the tissue cannot spontaneously relax back to the initial configuration~(Fig.~\ref{F_S8}c-h). Similarly to soap froths, which may be viewed as a passive analog of tissues, the spontaneous T1 transformations partly reduce the overall stress, yet the tissue still behaves like a solid for any chosen values of model parameters and therefore cells cannot rearrange so as to completely relax the stress. 
%

\begin{figure}[htb!]
	\centerline{\includegraphics{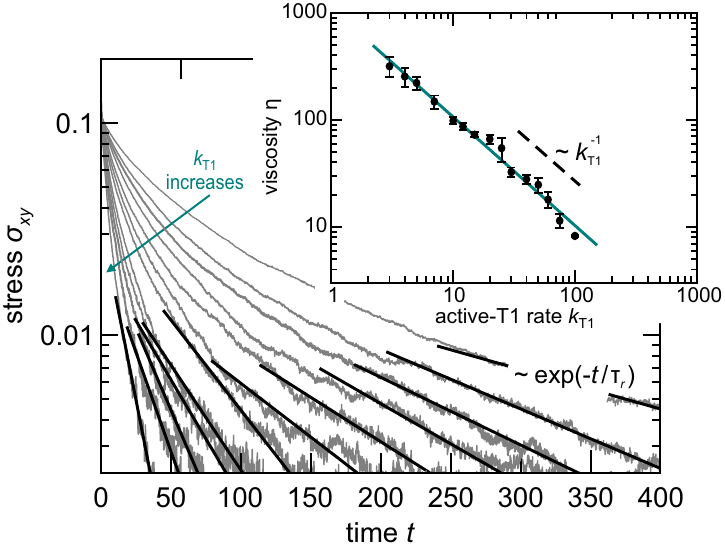}}
	\caption{\label{fig2}Stress-relaxation curves for $k_{\rm T1}=7,\>10,\>12,\>15,\>20,$ $\>25,\>30,\>40,\>50,\>60,\>75,$ and $100$ (top to bottom). The short-time elastic regime terminates at $t$ smaller than about 1/7,1/10,\ldots, 1/100, respectively. In the long-time viscous regime, stress decays exponentially; the relaxation time $\tau_r$ is obtained from fits (solid lines). Error bars were estimated by varying the fitting intervals. Inset: Viscosity $\eta$ vs.~active-T1 rate $k_{T1}$.}
\end{figure}

At long times, the stress decays to zero, which indicates a viscous response. In particular, 
$\sigma_{xy}\propto \exp{(-t/\tau_r)}$ where $\tau_r$ is the relaxation time, which is computed by fitting the stress-relaxation data averaged over 500 simulation runs (Fig.~\ref{fig2}). The obtained $\tau_r$ are then used to estimate the tissue viscosity $\eta$ using the relation $\eta=G\tau_r$ and study its dependence on the active-T1 rate. We find that $\eta$ rapidly decreases with $k_{\rm T1}$ and that 
\begin{equation}
	\label{eq8}
	\eta\propto k_{\rm T1}^{-1}\>
\end{equation}
as shown in the inset to Fig.~\ref{fig2}. 

To better understand this result, we develop a coarse-grained theory, using the approach described in Ref.~\cite{ranft10}. We write the stress as $\sigma_{ij}=\sigma\delta_{ij}+\widetilde\sigma_{ij}$, where $\sigma$ and $\widetilde\sigma_{ij}$ are its isotropic and traceless part, respectively. The rate of change of traceless stress $\widetilde\sigma_{ij}$ can be expressed as the sum of the elastic term and the source term due to active T1 transformations dressed by any subsequent rearrangements:
\begin{equation}
	\label{Eq_CG1}
	\frac{\textup D\widetilde\sigma_{ij}}{\textup Dt}=2G\widetilde v_{ij}+\widetilde\Pi_{ij}\>.
\end{equation}
Here $v_{ij}=(\partial_iv_j+\partial_jv_i)/2$ is the strain rate tensor and $\widetilde v_{ij}$ is its traceless part, $\widetilde\Pi_{ij}$ is the stress source, and $\textup D \widetilde\sigma_{ij}/\textup D t=\partial_t\widetilde\sigma_{ij}+v_k\partial_k\widetilde\sigma_{ij}+\omega_{ik}\widetilde\sigma_{kj}+\omega_{jk}\widetilde\sigma_{ik}$ is the convected corotational time derivative, $\omega_{ij}=(\partial_iv_j-\partial_jv_i)/2$ being the vorticity of the flow. 

When the individual dressed active T1 transformations do not interact with each other, the source term $\widetilde\Pi_{ij}=k_{\rm T1}\widetilde r_{ij}$, where $\widetilde r_{ij}$ denotes the stress generated by one such event. In absence of stress, the average $\left <\widetilde r_{ij}\right >$ must vanish on symmetry grounds both in the disordered and in the ordered tissues with infinite and six-fold rotational symmetry, respectively, even though each event provides a stress source. The presence of stress introduces a bias so that both $\left<\widetilde\sigma_{ij}\right >$ and $\left <\widetilde r_{ij}\right >$ are nonzero. As the dressed active T1 transformations reduce the stress,
\begin{equation}
	\left <\widetilde r_{ij}\right >=-\alpha\left <\widetilde\sigma_{ij}\right >\>,
\end{equation} 
where $\alpha$ is a dimensionless coefficient which measures the efficiency of the process. Thus $\left <\widetilde \Pi_{ij}\right> =-k_{\rm T1}\alpha\left <\widetilde\sigma_{ij}\right >$ and after averaging and rearranging Eq.~(\ref{Eq_CG1}) we obtain the Maxwell constitutive relation 
\begin{equation}
	\label{eqqqJ}
  	\left <\widetilde\sigma_{ij}\right >+\tau_M\left<\frac{\textup D\widetilde\sigma_{ij}}{\textup D t}\right >=2\eta\left <\widetilde v_{ij}\right >\>,
\end{equation}
where $\tau_M=1/(k_{\rm T1}\alpha)$ is the relaxation time and $\eta=\tau_M G$, which immediately gives $\eta=(G/\alpha)k_{\rm T1}^{-1}$ just like in the vertex model~[where the dimensionless coefficient $\alpha=5.8\cdot 10^{-4}$~(Fig.~\ref{fig2})]. This allows us to conclude that at long times where $\textup D\widetilde\sigma_{ij}/\textup Dt$ vanishes and $\left <\widetilde\sigma_{ij}\right >=2\eta\left <\widetilde v_{ij}\right >$, the epithelium can be treated as a viscous fluid. Equation~(\ref{eqqqJ}) introduces a single relaxation time. A more refined analysis leading to a distribution of these times is given in~Appendix~\ref{appD}; $\tau_M$ is related to this distribution via Eq.~(\ref{Eq_tempsM}).
\section{3D tissue shape}
\begin{figure}[htb!]
\centerline{\includegraphics[]{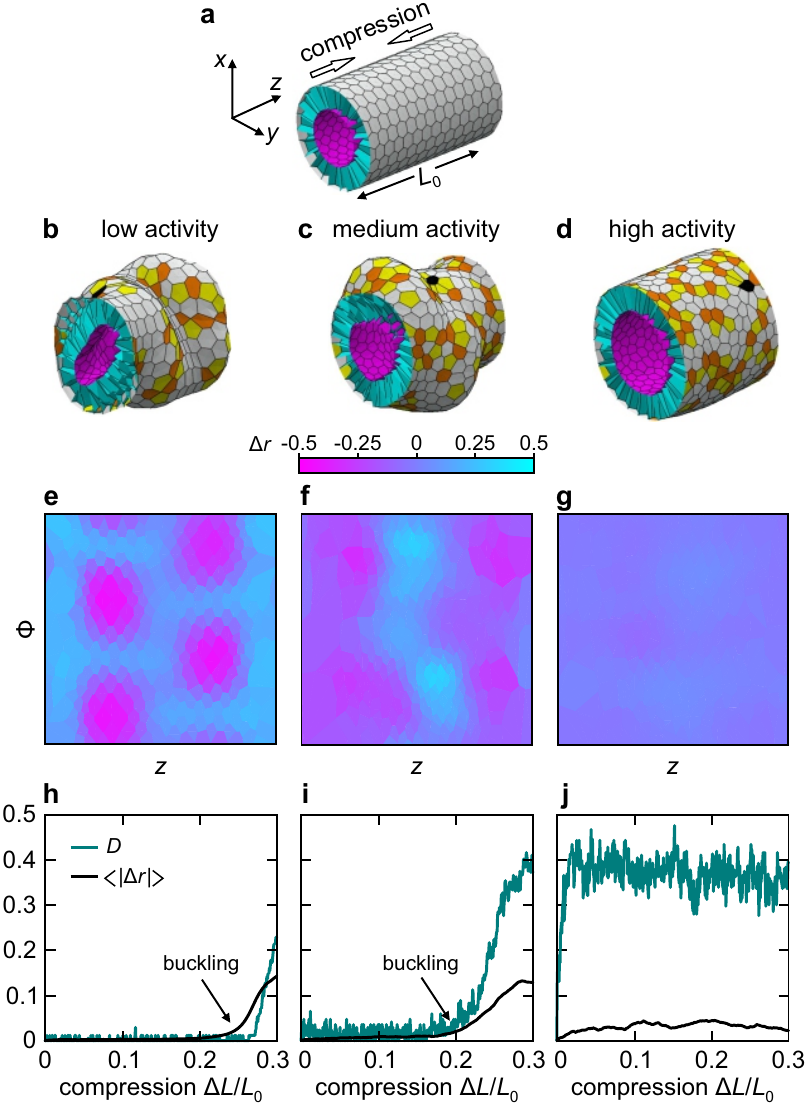}}
\caption{\label{fig4} Epithelial tube before (a) and after axial compression by $\Delta L/L_0=0.3$ at $k_{\rm T1}=3,\>20$, and $100$ (b, c, and d, respectively); the color-coded deformation patterns encoded by radial modulation $\Delta r$ are shown in panels e-g. Average magnitude of radial modulation $\left <\left |\Delta r\right |\right >$ and disorder $D$ vs.~relative compression at $k_{T1}=3, 20,$ and 100 (h, i, and j, respectively).}
\end{figure}

We now employ our model to demonstrate how the active T1 transformations affect the overall 3D shape of the tissue, examining the interplay of shape and structure in an axially compressed epithelial tube of $\mathcal C=336$ cells~(Fig.~\ref{fig4}a). Starting from an ordered state where all cells are 6-coordinated, we deform the tube at a constant compression rate $\Delta L/\tau_C=7\cdot 10^{-3}$, where $\tau_C=500$ is the time in which the tube is compressed by $30\%$, and at three different active-T1 rates $k_{\rm T1}$. At a small $k_{\rm T1}=3$, the tube is essentially elastic and buckles beyond the critical compressive strain~(Fig.~\ref{fig4}b and h). At $k_{\rm T1}=20$, the relaxation time $\tau_{M}$ is shorter but of the same order than $\tau_C$ and cells partially rearrange upon compression, which makes the tube smoother~(Fig.~\ref{fig4}c). The $k_{\rm T1}=100$ case (Fig.~\ref{fig4}d) corresponds to an extremely active tissue which readily flows while compressed, thereby avoiding buckling and merely transforming into a shorter tube with a larger diameter. Figures~\ref{fig4}e-g quantify the deviation of shapes in Figs.~\ref{fig4}b-d from the cylinder in terms of radial modulation $\Delta r=r/\left <r\right >-1$, where $r$ is the distance of a cell centroid from the axis and and $\left <r\right >$ is its average over all cells. Here $\Delta r$ is plotted in the $(z,\phi)$ plane, where $z$ is the lengthwise coordinate and $\phi$ is the polar angle. The evolution of these shapes is shown in Supplemental Movies M1-M3.

The in-plane structure of the tube is closely related to its 3D shape. In Figs.~\ref{fig4}h-j we plot the degree of disorder $D$ (defined by $1-\nu_6$ where $\nu_6$ is the frequency of 6-coordinated cells) and the average magnitude of $\Delta r$ against relative compression. In the $k_{\rm T1}=3$ and 20 tubes, the tissue remains ordered as long as it does not buckle. This is because the bound 4-cell pair of dislocations created by an active T1 transformation undergoes a reverse transformation before these cells can interact with other non-6-coordinated cells~(Figs.~\ref{fig4}h and i). Close to the buckling instability, more and more dislocations as well as disclinations survive so as to relax the stress and disorder is increased. In the fluidized tissue with $k_{\rm T1}=100$, the density of such defects quickly saturates and so does the value of $D$~(Fig.~\ref{fig4}j). This leads to a steady-state packing with a well-defined distribution of polygonal classes. 

\section{Conclusions}
The main novelty of our dynamical 3D vertex model of epithelia is the concept of tissue fluidization due to active noise at cell-cell junctions, which regulates the stress-relaxation timescale. This represents a new mechanism involved in the rigidity transition in tissues~\cite{bi16}, which does not rely on vanishing energy barriers and is thus independent on the choice of the work function. 

Our implementation of the active noise is based on exploration of the tissue energy landscape by forced neighbor exchanges. We showed that the geometry of the junctional network encodes some of the local features of this landscape. In particular, we discovered relations between the lengths of cell-cell junctions and energies involved in a T1 transformation. Using an elastic theory, we showed that these relations should not depend on the choice of the model-energy functional. 

Furthermore, devoid of simplifications immanent to its 2D analogs, our model can be used to explore the interplay of activity, in-plane structure, and 3D shape of tissues in various contexts, especially when generalized by including cell division and extrusion --- in order to, e.g., describe the form of villi and crypts in the intestinal epithelium (Sec.~II of Supplemental Material~\cite{SuppInf}). Important application will be a study of early embryonic development, say in {\sl Drosophila}, which will allow one to comprehensively explore the mechanics of the morphogenetic events involved. Here active T1 transformations may be biased by tissue in-plane polarization~\cite{Collinet15} and their rate (and thus the degree of fluidization) may vary between the different parts of the embryo. Finally, when further developed, our approach might be useful in finding robust ways to engineer tissues of different shapes and functionalities {\it in vitro}, e.g. organoids. 

\section{Acknowledgments}
The authors acknowledge the financial support from the Slovenian Research Agency (research core funding No. P1-0055), EMBO short-term fellowship, Slovene Human Resources Development and Scholarship Fund (Ad Futura grant), and the French Government Scholarship by Institut Fran\c{c}ais de Sloveni\'e. S.D. acknowledges the support of the Mechanobiology Institute, Singapore. This project has received funding from the European Unions Horizon 2020 research and innovation programme under the Marie Skłodowska-Curie ETN COLLDENSE, Grant Agreement No. 642774.
\appendix

\section{3D tension-based model\label{appA}}
We treat epithelial cell sheets as aggregates of incompressible cells carrying differential surface tensions at the apical, basal, and lateral cell sides ($\Gamma_a$, $\Gamma_b$, and $\Gamma_l$, respectively)~\cite{growthandform,steinberg63,harris76,derganc09}, such that the total mechanical energy of a cell reads 
\begin{equation}
	\label{E_S1}
	W=\Gamma_a A_a+\Gamma_bA_b+\frac{1}{2}\Gamma_lA_l\>,
\end{equation}
where $A_a$, $A_b$, and $A_l$ are the surface areas of the apical, basal, and lateral cell sides. Due to incompressibility, cell volume is fixed ($V=\textup{const.}$). For simplicity, we assume that all tensions are positive ($\Gamma_a,\>\Gamma_b,\>\Gamma_l>0$) which is true in the limit where the cortical tension is dominant over the adhesion strength. Let us describe cells as prisms with regular polygonal base. For a hexagonal base $A_a=A_b\equiv A$, $A_l=2h\sqrt{2\sqrt{3}A}$, and $V=Ah=\textup{const.}=1$, where $A$ is the surface area of the base and $h$ is the cell height. In dimensionless form where the dimensionless apical tension $\alpha=\Gamma_a/\Gamma_l$ and the dimensionless basal tension $\beta=\Gamma_b/\Gamma_l$, $a=A/V^{2/3}$, and $w=W/(V^{2/3}\Gamma_l)$, the total energy [Eq.~(\ref{E_S1})] of a cell reads
\begin{equation}
	\label{E_S2}
	w=\left ( \alpha+\beta\right )  a+\frac{\sqrt{2\sqrt{3}}}{\sqrt{a}}\>.
\end{equation}
The equilibrium surface area of the cell base $a_0$ is calculated from the force balance equation \Big ($\partial w/\partial a=\alpha+\beta-\sqrt{\sqrt{3}/2a^3}=0$\Big ) and reads
\begin{equation}
	\label{E_S3}
	a_0=\left (\frac{\sqrt{3}}{2}\right )^{1/3}(\alpha+\beta)^{-2/3}\>,
\end{equation}
whereas the equilibrium cell height $h_0$ is calculated from the fixed-volume constraint $h_0=1/a_0$:
\begin{equation}
	\label{E_S4}
	h_0=\left (\frac{\sqrt{3}}{2}\right )^{-1/3}(\alpha+\beta)^{2/3}\>.
\end{equation} 
By combining Eqs.~(\ref{E_S2})~and~(\ref{E_S3}), we obtain the equilibrium cell energy $w_0=3\left [\sqrt{3}(\alpha+\beta)/2\right ]^{1/3}$. In a flat geometry, the ground state corresponds to a tissue where cells of height $h_0$ and cell-base area $a_0$ are packed in a regular hexagonal lattice with 6-coordinated cells.

The most basic classification of epithelial monolayers relies on their width-to-height ratio $\Phi=d/h$, where $d$ is the cell diameter. The preferred cell width-to-height ratio, defined as $\Phi_0=d_0/h_0$ where $d_0=\sqrt{2a_0/\sqrt{3}}$, reads
\begin{equation}
	\label{E_S6}
	\Phi_0=\frac{1}{\alpha+\beta}\>.
\end{equation}
For $\alpha+\beta<1$ the tissue is squamous ($d/h>1$), for $\alpha+\beta>1$ the tissue is columnar ($d/h<1$), whereas for $\alpha+\beta\approx 1$ the tissue is cuboidal ($d/h\approx 1$). 

Generally, tissue can spontaneously fold due to apico-basal polarity~\cite{krajnc15}, which can be modeled by a non-zero apico-basal differential tension ($\alpha-\beta\neq0$). Here we do not consider such a case and thus we assume that $\alpha=\beta\equiv\gamma$. 
\section{Collapse of data for $\delta w$ and $w-w'$\label{appB}}
Let us denote the surface area of the midplane of cell $i$ by $a_i\approx\left[a_a^{(i)}+a_b^{(i)}\right]/2$, such that the total energy of the system~[Eq.~(\ref{eq1})] simplifies to 
\begin{align}
	\label{Renergy}
	w&=\sum_{i=1}^{\mathcal C}\left [2\gamma a_i+\frac{1}{2}a_l^{(i)}\right ]\>,
\end{align}
where $a_l^{(i)}\approx p_ih_i$ is the total lateral area of cell $i$ ($p_i$ being the perimeter of the midplane). Assuming that cell height is uniform across the tissue (i.e., $h_i=h_0=2\cdot 3^{-1/6}\gamma^{2/3}$) and that the volumes of all cells are fixed, we can write $a_i\approx 1/h_0$ for all $i$ so that the total energy
\begin{equation}
	w=\frac{2\mathcal C\gamma}{h_0}+\frac{h_0}{2}\sum_{i=1}^{\mathcal C}p_i\>.
\end{equation}
This can be further simplified by rewriting the sum over all cells as a sum over all edges such that
\begin{equation}
	w=\frac{2\mathcal C\gamma}{h_0}+h_0\sum_{i=1}^{\mathcal E}l_i=\textup{const.}+h_0\sum_{i=1}^{\mathcal E}l_i\>.
\end{equation}
This result suggests that it might be possible to collapse the data for $\delta w$ and $w-w'$ [Eqs.~(\ref{Eqbarrier}) and (\ref{Eqgain}), respectively] by rescaling the energy by $h_0l_0$ and the edge length by $l_0$, where $l_0=3^{-2/3}\gamma^{-1/3}$ is the lattice constant of the corresponding ground-state honeycomb lattice. Thus, the rescaled energy barrier 
\begin{equation}
	\frac{\delta w}{h_0l_0}=R\left (\frac{l'}{l_0}\right )^2\>,
\end{equation}
whereas the rescaled energy gain
\begin{equation}
	\frac{w'-w}{h_0l_0}=Q\left (\frac{l'}{l_0}-\frac{l^*}{l_0}\right )\>.
	\label{druga}
\end{equation}
\section{Comparison to APE model and 2D soap froth\label{appC}}
To test whether the obtained relationships $\delta w(l')$ and $(w-w')(l')$ hold for other tissue model-energy functionals as well, we perform the analysis of energy barriers and energy gains for two other mechanical models: 2D foam model and the APE model. We consider a planar polygonal tiling of straight edges. The model dimensionless energy in the 2D foam
\begin{equation}
	w_{\rm 2Dfoam}=\sum_{i=1}^{\mathcal E}l_i+k_A\sum_{i=1}^{\mathcal C}\left (a_i-1\right )^2\>,
\end{equation}
whereas in the APE model
\begin{equation}
	w_{\rm APE}=\sum_{i=1}^{\mathcal C}\left (p_i-p_0\right )^2+k_A\sum_{i=1}^{\mathcal C}\left (a_i-1\right )^2\>,
\end{equation}
where $p_0$ is the preferred perimeter of cells. In both models we assume that the cells are nearly incompressible and set $k_A=100$. 
\begin{figure}[htb!]
	\begin{center}
	\includegraphics[]{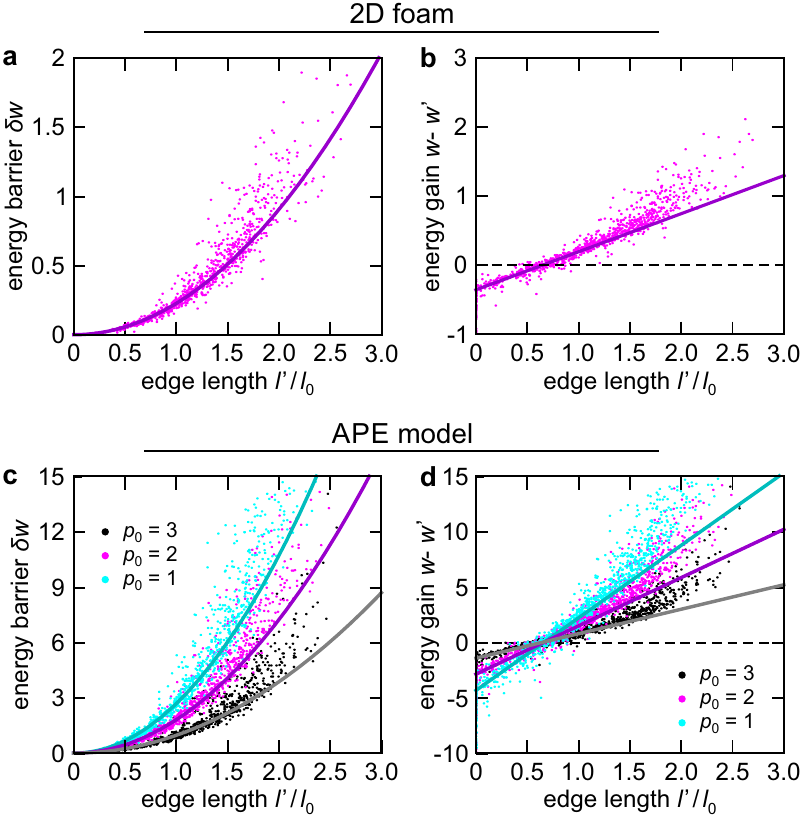}
	\caption{\label{F_S5}Energy barrier $\delta w/w_0$ vs.~edge length $l'/l_0$~(a, c) and energy gain $(w-w')/w_0$ vs.~edge length $l'/l_0$~(b, d) for the 2D-foam model and the 2D APE model, respectively. Dots are simulation data and the solid lines are Eqs.~(\ref{Eqbarrier}) and (\ref{Eqgain}), respectively. The threshold value $l^*/l_0\approx 0.64$ and 0.66~(panels b and d, respectively).}
	\end{center}
\end{figure}

We construct a random tiling of cells using the Voronoi construction and perform T1 transformation on all edges just like in the 3D tension-based model~(Fig.~\ref{F_S4}). We measure $\delta w$ and $w-w'$ and plot them versus the initial edge rest length $l'$~(Fig.~\ref{F_S5}). The 2D-foam model has no free parameters, whereas for the APE model, we plot $\delta w$ and $w-w'$ for three values of the preferred perimeter $p_0=1,2,$ and $3$ (all corresponding to the solid-tissue regime). We find that just like in the 3D tension-based model, $\delta w\propto l'^2$ and $w-w'\propto l'-l^*$. Furthermore, the threshold length in both 2D foam and in 2D APE model agrees very well with that of the 3D tension-based model. In particular, we find $l^*_{\rm 2Dfoam}\approx0.64l_0$ and $l^*_{\rm APE}\approx0.66l_0$, where $l_0=\sqrt{2}/3^{3/4}$ is the lattice constant of the lattice of unit-area regular hexagons~(Fig.~\ref{F_S5}b and d).
\section{Non-exponential stress relaxation\label{appD}}
We start from the relation between the stress derivative, the strain rate and the stress sources due to imposed T1 transformations and subsequent relaxation as introduced in the main text:
 \begin{equation}
	\label{Eq_CGS1}
	\frac{\textup D\widetilde\sigma_{ij}}{\textup Dt}=2G\widetilde v_{ij}+\widetilde\Pi_{ij}\>.
\end{equation}
The stress source $\widetilde\Pi_{ij}=k_{\rm T1}\left <\widetilde r_{ij}\right >$ is the sum of all stresses corresponding to pairs of forced T1/relaxations events. Each pair of events has its own time scale and amplitude and the relation $
	\left <\widetilde r_{ij}\right >=-\alpha \left <\widetilde\sigma_{ij}\right > 
$ used in a mean-field approach, valid only at long times. Here we elaborate this simple mean-field description by assuming that each pair of events is characterized by an initial edge length $l'$. (In principle, it also depends on the the orientation of the edge in question, cell connectivity etc.)  Keeping simply the edge length dependence is again a simplification but it allows us to understand the non-exponential relaxation observed in the simulations in a simple way. 

The relation between the average source stress and that of events with initial edge length $l'$ is given by
\begin{equation}
\label{Eq_stressl}
\left <\widetilde r_{ij}\right >=\int_{0}^{l'_{\rm max}} \textup dl' P_{l'}\left <\widetilde r_{ij}^{l'}\right > \>,
\end{equation}
where $P_{l}$ is the edge length probability distribution and $\left <\widetilde r_{ij}^{l'}\right >$ is the average stress source corresponding to events with initial edge length $l'$. Here $l'_{\rm max}$ is the maximal possible edge length. The dynamics of such events can be characterized by a time $\tau_{l'}$ and an amplitude $\chi_{l'}$ such that in the linear regime:
\begin{equation} 
\label{Eq_pil}
\tau_{l'} \frac{\partial \left <\widetilde r_{ij}^{l'}\right >}{\partial t}+\left <\widetilde r_{ij}^{l'}\right >=-\chi_{l'}\left <\widetilde \sigma_{ij}\right >
\end{equation}
or after integration
\begin{equation}
\label{Eq_pil-int}
\left <\widetilde r_{ij}^{l'}(t)\right >=-\frac{\chi_{l'}}{\tau_{l'} }\int_{0}^{t} \exp{\left (-\frac{t-s}{\tau_{l'}}\right )} \left <\widetilde \sigma_{ij}(s)\right >\textup ds
\end{equation}
Now the stress evolution equation reads:
\begin{eqnarray}
	\label{Eq_ImprovedMaxwell}
	\frac{\textup D\widetilde\sigma_{ij}}{\textup Dt}&=&2G\widetilde v_{ij}\\
	&& -k_{\rm T1}\int_{0}^{t} \textup ds \int_{0}^{l'_{\rm max}} \textup dl' \frac{P_{l'}^{ss}\chi_{l'}}{\tau_{l'}}\exp{\left (-\frac{t-s}{\tau_{l'}}\right )}\widetilde\sigma_{ij}(s)\>.\nonumber
\end{eqnarray}
For long enough times or for slow enough stress variations, one recovers the Maxwell relation in the main text [Eq.~(7) of the main text] with:
\begin{equation}
\label{Eq_tempsM}
\tau_{M}^{-1}=k_{\rm T1}\int_{0}^{l'_{\rm max}} \textup dl' P_{l'}\chi_{l'}
\end{equation}
However, note that Eq.~(\ref{Eq_ImprovedMaxwell}) involves a continuum of time scales as observed in the simulation. This is particularly clear in the Fourier space, where the linearized relation between stress and strain now reads:
\begin{equation}
\label{Eq_Fourier}
\left (i\omega+k_{\rm T1}\int_{0}^{l'_{\rm max}} \frac{\textup dl' P_{l'}\chi_{l'}}{1+\omega^{2}\tau_{l'}^{2}}\right )\left <\widetilde \sigma_{ij}(\omega)\right >=2G\widetilde v_{ij}(\omega)\> .
\end{equation}

\end{document}



\title{Fluidization of sheet-like tissues by active cell rearrangements \break Supplemental Material}
\author{Matej Krajnc}
\affiliation{Lewis-Sigler Institute for Integrative Genomics, Princeton University, Princeton, USA}
\affiliation{Jo\v zef Stefan Institute, Ljubljana, Slovenia}
\author{Sabyasachi Dasgupta}
\affiliation{Department of Physics, University of Toronto, 60 St.~George St., Toronto, ON M5S 1A7, Canada}
\affiliation{Mechanobiology Institute, National University of Singapore, 5A Engineering Drive 1, Singapore 117411, Singapore}
\author{Primo\v z Ziherl}
\affiliation{Jo\v zef Stefan Institute, Ljubljana, Slovenia}
\affiliation{Faculty of Mathematics and Physics, University of Ljubljana, Ljubljana, Slovenia}
\author{Jacques Prost}
\affiliation{Mechanobiology Institute, National University of Singapore, Singapore}
\affiliation{Institute Curie, Paris, France}


\maketitle

\setcounter{figure}{0}
\renewcommand{\thefigure}{S\arabic{figure}}

\setcounter{equation}{0}
\renewcommand{\theequation}{S\arabic{equation}}
%
\section{3D vertex model}
Cells in an epithelial sheet are described by their 3D shape as well as by their polygonal cell packing characterized by the network of cell-cell junctions. In order to relax the local stress, cells change their shape and they reorganize within the monolayer. To capture this dynamics, we use a 3D vertex model where cells are represented by vertices and their dynamics is simulated by integrating the equation of motion for all vertices. This is complemented by topological transformations of the edge network leading to local rearrangements within the tissue network which then allows the flow of individual cells within the monolayer. 
\subsection{Implementation}
Within a 3D vertex model, the basal side of a single-cell-thick epithelium is represented by $\mathcal V$ vertices connected by $\mathcal E$ edges. A generally non-flat polygonal network of edges contains $\mathcal C$ polygonal cell bases representing the basal side of the tissue (i.e.~the basal network). On a torus (i.e. periodic boundary conditions in the direction of two spatial coordinates, say $x$ and $y$), $\mathcal V$, $\mathcal E$, and $\mathcal C$ are related by the Euler formula:
\begin{equation}
	\label{E_S8}
	0=\mathcal V-\mathcal E+\mathcal C\>.
\end{equation}
We assume that the apical network has the same connectivity as the basal network so that in terms of topology, it is the exact copy of the basal network. Of course, each vertex in the apical network is allowed to have all three spatial coordinates $x$, $y$, and $z$ different from those of its basal counterpart. Next, apical and basal networks are connected by lateral edges joining the basal vertices with their corresponding apical vertices. The 3D network of vertices and edges represents the tissue skeleton (i.e.~cell outlines) and the apical, basal, and lateral polygonal cell sides are defined as triangulated surfaces.

A 3D tissue is uniquely defined by the basal network and normal vectors to all basal vertices, which together specify both the direction as well as the distance between the basal vertex and its apical counterpart. Working only with these elements is an efficient way of dealing with a 3D vertex model of a single-cell-thick epithelium. Below we list the key technical details of the implementation.

\subsection*{Basal network}
\begin{figure}[!htb]
	\begin{center}
	\includegraphics[]{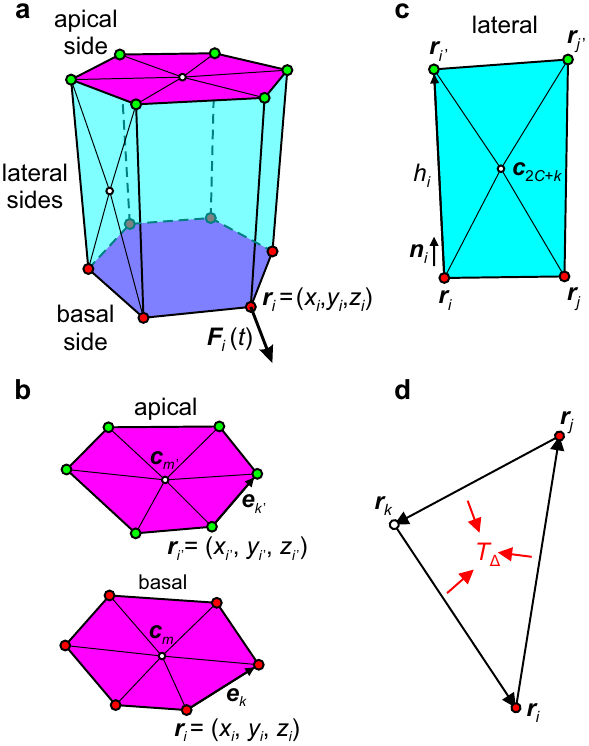}
	\caption{\label{F_S2}
	(a)~Schematic of an epithelial cell. Within the 3D vertex model, cells are represented by basal and apical vertices (red and green circles, respectively), and forces exerted on vertices $\boldsymbol F_i$ are calculated from their positions $\boldsymbol r_i$. (b)~The apical, basal and lateral surfaces are triangulated using passive vertices placed at the centers of cell sides (white circles). (c)~Each vertex is assigned a unit normal vector $\boldsymbol n_i$ and length of the lateral edge associated with this vertex $h_i$. (d)~Surface tension $T_\triangle$ acting on a triangular element defined by vertices $\boldsymbol r_i$, $\boldsymbol r_j$, and $\boldsymbol r_k$.}
	\end{center}
\end{figure}

The positions of basal vertices
\begin{equation}
	\label{E_S9}
	\boldsymbol r_i=(x_i,y_i,z_i)\>
\end{equation}
for $i=1,2,\ldots\>\mathcal V$~(Fig.~\ref{F_S2}a). Vertex $i$ is connected to vertex~$j$ with an oriented basal edge $k$, which is conveniently defined as
\begin{equation}
	\label{E_S10}
	\boldsymbol e_k=\boldsymbol r_i-\boldsymbol r_j\>
\end{equation}
for $k=1,2,\ldots\>\mathcal E$ and $i,j\neq i\in\{1,2,\ldots\>\mathcal V\}$~(Fig.~\ref{F_S2}b). The polygonal cell base $m$ is defined as $\boldsymbol p_m$, a list of oriented identification numbers of edges $k_s^{(m)}$ which enclose the base in counterclockwise direction:
\begin{equation}
	\label{E_S11}
	\boldsymbol p_m=\left (\sigma_{1}^{(m)}k_{1}^{(m)},\sigma_{2}^{(m)}k_{2}^{(m)},\sigma_{3}^{(m)}k_{3}^{(m)},\ldots\>\sigma_{S_m}^{(m)}k_{S_m}^{(m)}\right )\>
\end{equation} 
for $m=1,2,\ldots\>\mathcal C$. Here $\sigma_s=\pm 1$ specifies the orientation of an edge within cell base $m$ and $s=1,2,\ldots\>S_m$, where $S_m\geq 3$ is the number of edges of polygon $m$. Each vertex~$i$ is also assigned a unit vector $\boldsymbol n_i=\left (n_x^{(i)},n_y^{(i)},n_z^{(i)}\right )$ and a scalar $h_i$ which carry the information of the direction and the distance between basal vertex $i$ and its apical counterpart $i'$, respectively~(Fig.~\ref{F_S2}c).
%
\subsection*{Apical network}
Geometric elements belonging to the apical network all have their counterparts in the basal network. We therefore denote their identification numbers using the same symbol as for the basal network but adding the prime ($'$). For example, edge $\boldsymbol e_{k'}$ is the apical counterpart of the edge $\boldsymbol e_{k}$ in the basal network. 

The positions of apical vertices are given as 
\begin{equation}
	\label{E_S12}
	\boldsymbol r_{i'}=\boldsymbol r_i+h_i\boldsymbol n_i\>,
\end{equation}
where $i'=\mathcal V+i$ ($i=1,2,\ldots\>\mathcal V$)~(Figs.~\ref{F_S2}a-c). The oriented apical edges are derived from the positions of apical vertices $\boldsymbol r_{i'}$ as follows:
\begin{equation}
	\label{E_S13}
	\boldsymbol e_{k'}=\boldsymbol r_{i'}-\boldsymbol r_{j'}\>,
\end{equation}
where a pair $(i',j')$ corresponds to a pair $(i,j)$ defining an edge $k$, $k'=\mathcal E+k$ ($k=1,2,\ldots\>\mathcal E$), and $j'=\mathcal V+j$ ($j\neq i\in\{1,2,\ldots\>\mathcal V\}$)~(Fig.~\ref{F_S2}b). A polygonal cell apex $m'$ is defined by $\boldsymbol p_{m'}$, a list of oriented identification numbers of edges $k_s^{(m')}$ which enclose the apex in counterclockwise direction:
\begin{equation}
	\label{E_S14}
	\boldsymbol p_{m'}=\left (\sigma_{1}^{(m')}k_{1}^{(m')},\sigma_{2}^{(m')}k_{2}^{(m')},\sigma_{3}^{(m')}k_{3}^{(m')},\ldots\>\sigma_{S_{m'}}^{(m')}k_{S_{m'}}^{(m')}\right )\>.
\end{equation}
		Here $m'=\mathcal C+m$ ($m=1,2,\ldots\>\mathcal C$). The list $\boldsymbol p_{m'}$ is derived from the list $\boldsymbol p_{m}$ as follows:
\begin{subequations}
	\label{E_S15}
	\begin{equation}
		\sigma_{s}^{(m')}=\sigma_{s}^{(m)}\>,
	\end{equation}
	\begin{equation}
		k_{s}^{(m')}=\left (k_{s}^{(m)}\right )'\>,
	\end{equation}
\end{subequations} 
for $s=1,2,\ldots\>S_{m'}$. The topologies of the basal and the apical network are by assumption the same and thus
\begin{equation}
	\label{E_S16}
	S_m=S_{m'}\>.
\end{equation}
%
\subsection*{Facets and cells}
Generally, the vertices of a given polygon are non-coplanar and therefore the surface areas of apical, basal, and lateral sides cannot be well defined. To this end, we triangulate all surfaces by introducing center vertices of cell sides such that each triangle of a triangulated polygon consists of two consecutive vertices from the basal/apical network and the center vertex~(Figs.~\ref{F_S2}b and c). 

Center vertices are passive, i.e.~they are always positioned at the center of a given cell side and their positions are not controlled by the equation of motion. The positions of basal and apical cell side centers are calculated using
\begin{subequations}
	\label{E_S18}
	\begin{equation}
		\boldsymbol c_m=\frac{1}{S_m}\sum_{i\in \textup{polygon }m} \boldsymbol r_i\>,
	\end{equation}
{\rm and} \hfill		
	\begin{equation}
		\boldsymbol c_{m'}=\frac{1}{S_{m'}}\sum_{i\in \textup{polygon }m'}\boldsymbol r_{i'}\>,
	\end{equation}	
\end{subequations}
whereas the positions of lateral side centers are given by
\begin{equation}
	\label{E_S19}
	\boldsymbol c_{2C+k}=\frac{\boldsymbol r_{i}+\boldsymbol r_{i'}+\boldsymbol r_{j}+\boldsymbol r_{j'}}{4}\>,
\end{equation}
where $i$ and $j$ are indices of the starting and the ending vertex of an edge $k$. Positions of center vertices $\boldsymbol c$ are updated at each time step. 

Finally, cell $m$ is defined by $S_m$ oriented triangles forming the basal side $m$, $S_{m'}=S_m$ triangles forming the apical side $m'$ and $4S_m$ triangles forming the corresponding lateral sides~(Figs.~\ref{F_S2}b and c).
\subsection{T1 topological transformation}
Within our implementation, topological transformations are effectively 2D. A topological transformation is applied by changing the connectivity of the basal network stored in $\boldsymbol p_m$~[Eq.~(\ref{E_S11})], and all other elements such as the connectivity of the apical network, lateral sides etc.,~are then updated accordingly. 
\begin{figure}[!htb]
	\begin{center}
	\includegraphics[]{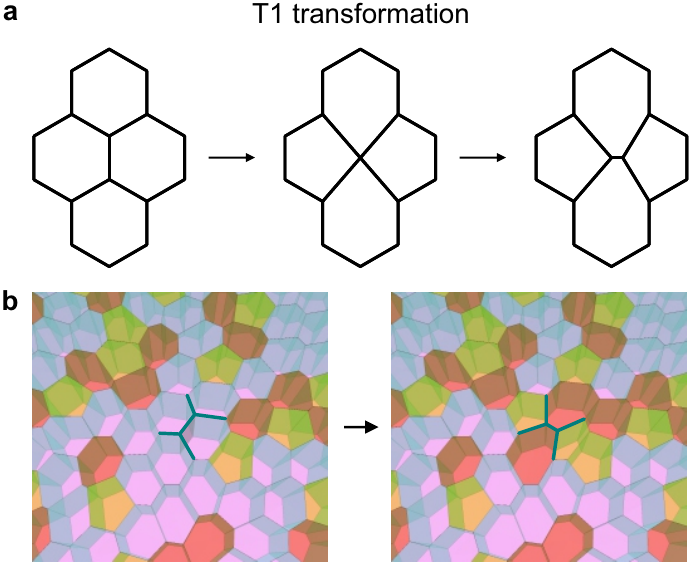}
	\caption{\label{F_S3}
	(a)~T1 transformation on a four-cell neighborhood. Two three-way vertices are merged to form a four-way vertex. This is followed by the reduction of the valence of the vertex such that a new edge (and a new vertex) is formed between cells that initially did not share edges. (b)~Snapshots of simulation showing T1 transformation. The color coding corresponds to different polygon classes (the semitransparent shading of the apical sides of 5-, 6-, and 7-coordinated cells is yellow, white, and brown, respectively).}
	\end{center}
\end{figure}

In tissues, cells rearrange so as to relax the local stress. Rearrangement occurs in multiple steps by successive edge swaps commonly referred to as T1 topological transformations~(Fig.~\ref{F_S3}). In many aspects, T1 transformation is the most important topological transformation in the dynamics of cellular networks and involves a four-cell neighborhood in which a pair of cells (cells 1 and 2) share an edge, while the other pair (cells 3 and 4) is tethered by the same edge~(Fig.~\ref{F_S3}a). In the first step, the edge shared by cells 1 and 2 is deleted and the associated vertices are merged to form a four-way vertex. In the second step, a new edge is formed between cells 3 and 4, which initially did not share any edges. T1 transformation conserves the number of vertices $\mathcal V$, edges $\mathcal E$, and cells~$\mathcal C$. 

In our simulations, a T1 transformation takes place as follows. At a time $t$, a chosen edge between cells 1 and 2 is deleted and the associated vertices are merged to form a four-way vertex. The structure then evolves in time ($t\to t+\delta t$), where $\delta t$ is the simulation time step, and the volume-restoring motion is applied to adjust the cell volumes which were slightly changed due to merged vertices~\cite{brakke}. The four-way vertex is short-lived and decays after a time of the order of $10\delta t$ after which a new (short) edge between cells 3 and 4 is formed. This is illustrated in the simulation snapshots in Fig.~\ref{F_S3}b. 
\subsection{Vertex dynamics}
In the vertex model, the tissue is represented by vertices which are moved during simulation so as to minimize the total mechanical energy of the system. The vertices are connected by edges to define prismatic cells with polygonal bases, which adhere to each other by sharing facets. We assume that all cells have identical properties and (in general) allow for different surface tensions on the apical, basal, and lateral cell sides ($\Gamma_a$, $\Gamma_b$, and $\Gamma_l$, respectively) such that the total energy of the tissue can be written as a sum over all cells:
\begin{equation}
	\label{E_S20}
	W=\sum_{i=1}^{\mathcal C} \left [\Gamma_a A_a^{(i)}+\Gamma_b A_b^{(i)}+\frac{1}{2}\Gamma_l A_l^{(i)}\right ]\>,
\end{equation}
where $A_a^{(i)}$, $A_b^{(i)}$, and $A_l^{(i)}$ are surface areas of the apical, basal, and lateral sides of cell $i$, respectively, and $\Gamma_a$, $\Gamma_b$, and $\Gamma_l$ are all assumed positive ($\Gamma_a,\Gamma_b,\Gamma_l>0$). The factor $1/2$ in front of the lateral term is needed because the lateral sides are shared by two cells. We treat cells as incompressible so that their volume is kept fixed ($V_i=V=\textup{const.}$).

In vertex models, the equation of motion of cells is given implicitly by the equation of motion of vertices. We assume that the inertial term is small compared to the friction term and thus vertices undergo first-order dynamics given by the overdamped equation of motion
\begin{equation}
	\label{E_S21}
	\eta_i\frac{\textup d\boldsymbol r_i(t)}{\textup d t}=\boldsymbol F_i(t)\>.
\end{equation}
Here $\boldsymbol r_i(t)=(x_i,y_i,z_i)$ is the position of vertex $i$ at a time $t$, $\boldsymbol F_i(t)$ is the total force exerted on vertex $i$ at a time $t$, and $\eta_i$ denotes the friction drag coefficient, i.e.~inverse vertex mobility $\mu_i$ ($\eta_i=1/\mu_i$). At each time step, the positions of vertices are updated as
\begin{equation}
	\label{E_S22}
	\boldsymbol r_i(t+\textup dt)=\boldsymbol r_i(t)+\frac{1}{\eta_i}\boldsymbol F_i(t)\textup dt\>,
\end{equation}
the forces exerted on vertices $\boldsymbol F_i=-\boldsymbol\nabla_i W(\boldsymbol r)$ where $\boldsymbol\nabla_i=\left (\partial/\partial x_i,\partial/\partial y_i,\partial/\partial z_i\right )$ and $\boldsymbol r=(\boldsymbol r_1, \boldsymbol r_2, \boldsymbol r_3, \ldots)$. The time scale of the dynamics is obtained by expressing the energy~[Eq.~(\ref{E_S20})] and the equation of motion~[Eq.~(\ref{E_S21})] in a dimensionless form. All lengths are measured in units of $V^{1/3}$, the energy in units of $\Gamma_l V^{2/3}$, and surface tensions in units of lateral surface tension $\Gamma_l$. Assuming $\Gamma_a/\Gamma_l=\Gamma_b/\Gamma_l=\gamma$, the dimensionless energy reads
\begin{equation}
	\label{E_S23}
	w=\sum_{i=1}^{\mathcal C} \left [\gamma\left (a_a^{(i)}+a_b^{(i)}\right )+\frac{1}{2}a_l^{(i)}\right ]\>.
\end{equation}
We assume that all vertices have the same mobility ($\mu_i=\mu_0$ for all $i$); in this case, the time scale is given by $\tau=1/(\mu_0\Gamma_l)$.
%
\subsection*{Force calculation}
In our surface-tension-based model, force calculation is fairly simple, because it only requires calculating gradients of the surface areas of triangular elements. The force exerted on vertex $i$
\begin{equation}
	\boldsymbol F_i=-\sum_{\triangle, i\in\triangle} T_\triangle \nabla_i a_\triangle\>,
\end{equation}
where $T_\triangle\in\{1,\gamma\}$ and $a_\triangle$ are the surface tension prescribed to triangle $\triangle$ and its surface area, respectively, and the sum goes over all triangles containing vertex $i$. Let us consider a triangular surface element $\triangle$ (i.e.~facet) defined by the vertices $\boldsymbol r_i$, $\boldsymbol r_j$, and $\boldsymbol r_k$ in counterclockwise order~(Fig.~\ref{F_S2}d). The gradient of the surface area of triangular facet $\triangle$ is calculated as
\begin{equation}
	\label{E_S25}
	\nabla_ia_{\triangle}=\frac{1}{2}\nabla_i\left |(\boldsymbol r_i-\boldsymbol r_k)\times(\boldsymbol r_j-\boldsymbol r_k)\right |\>.
\end{equation}
The fixed-volume constraint on all cells is satisfied by the volume-restoring motion and projection of forces to the subspace of volume-preserving motions as described in Ref.~\cite{brakke}.
\section{Villi and crypts}
The epithelia that cover organs in the digestive tract often assume a corrugated shape so as to maximize the effective surface area for the exchange of nutrients. The relationship between shape and function is especially prominent in finger-like structures found, e.g., in the intestine and referred to as villi and crypts. It is well known that in the crypts, cells rapidly divide and then they migrate toward the tip of the villi where they are extruded from the tissue. The exact shape of these structures is determined by curvature-dependent division/extrusion rates as well as by the interaction of the tissue with the underlying stroma and the basement membrane. The apico-basal polarity, which can be modelled by the apico-basal differential tension, may also play an important role.

\begin{figure}[htb!]
	\begin{center}
	\includegraphics[width=88mm]{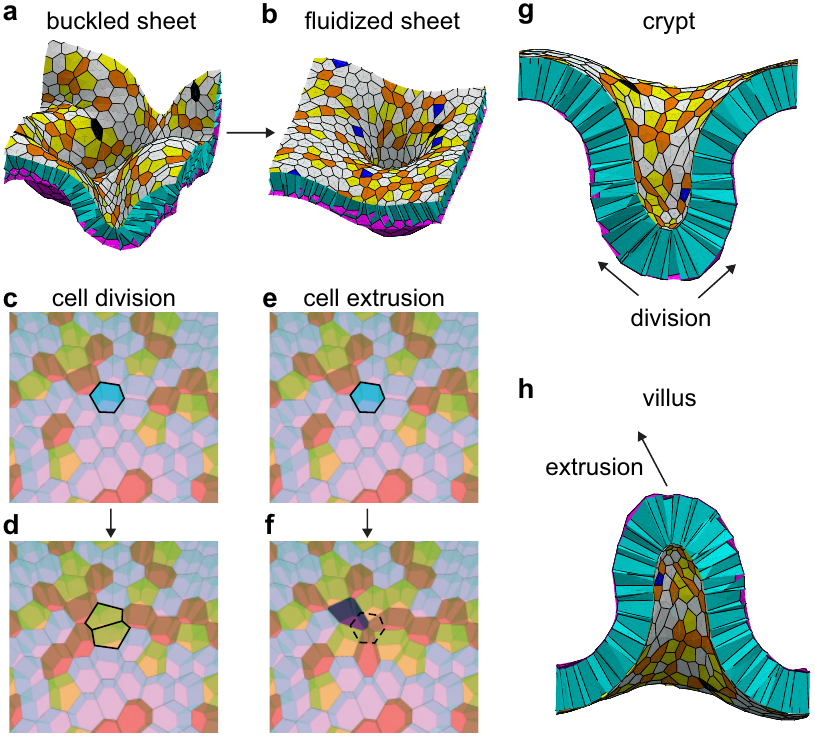}
	\caption{\label{F_S10}
	(a)~A flat cell sheet buckles when isotropically compressed. The deformation smoothens out by fluidization of the tissue using spontaneous and active T1 transformations~(panel b). (c-f)~Snapshots of a simulation showing cell division~(panels c and d) and cell extrusion~(panels e and f). 
	The finger-like protrusion obtained by fluidization of a buckled sheet can be seen as a proxy for villi and crypts (panels g and h, respectively). The color coding corresponds to different polygon classes~(same as in Fig.~\ref{F_S3}).}
	\end{center}
\end{figure}

To properly address the cellular dynamics and the overall shape of villi and crypts, the 3D vertex model presented in the main text needs to be generalized by including cell division and extrusion as additional topological transformations. In a preliminary step, we here do not consider division and extrusion but instead take a buckled sheet as a proxy for villi. We start with a flat epithelial sheet with periodic boundary conditions. Instead of shearing the simulation box, we make it smaller by decreasing the dimensions in both $x$- and $y$- directions. By minimizing the energy at the fixed topology, we obtain a somewhat irregularly buckled sheet~(Fig.~\ref{F_S10}a).  

After fluidizing the tissue by both spontaneous and active T1 transformations, the deformation profile smoothens out and the final shape has a single finger-like protrusion. Since the apical and the basal cell side are equivalent as far as surface tensions are concerned, the obtained steady-state shape can be viewed either as a crypt or as a villus~(Fig.~\ref{F_S10}g and h, respectively). Our future work will be dedicated to adding a curvature-dependent probability for division and extrusion~(Fig.~\ref{F_S10}c-f and Ref.~\cite{hannezo11}) focusing on how the division/extrusion mechanism interplays with the apico-basal differential surface tension. Using this model, we expect to obtain a steady-state flow of cells from crypts to the tip of the villi, allowing us to understand what determines the steady-state 3D shape of these structures.

An active tissue at a fixed $k_{\rm T1}$ never reaches the ground state, but stays in an excited state with a well-defined in-plane structure characterized by the distribution of polygonal classes. Nevertheless, our approach can be generalized to an energy-minimizing scheme based on the idea of simulated annealing, where the rate of active T1 transformations $k_{\rm T1}$ is gradually decreased during simulation. Such a dynamics should result in a state close to the global energy minimum in terms of both 3D shape and in-plane structure of the tissue.
